# A magnonic-optoelectronic reservoir for physical reservoir computing

Alexey B. Ustinov[1], Ivan Y. Tatsenko[1], Andrei A. Nikitin[1], Mikhail P. Kostylev[2]

1. Department of Physical Electronics and Technology, St. Petersburg Electrotechnical University, St. Petersburg 197022, Russia

2. Department of Physics, University of Western Australia, Crawley, Western Australia 6009, Australia

**Corresponding author**
Ivan Y. Tatsenko
Department of Physical Electronics and Technology
St. Petersburg Electrotechnical University
St. Petersburg 197022, Russia
E-mail: abitur.tatsenko@mail.ru

**Abstract**

Physical reservoir computing is a promising approach for fast and energy efficient computer vision, natural language processing, and general pattern recognition. This work presents a physical reservoir based on magnonic-optoelectronic oscillator (MOEO). An important feature of the device's optical path is the use of a fiber-optic delay line as a short-term-memory element. The microwave path is responsible for nonlinear mapping of input data to a higher-dimensional space. The strong four-wave nonlinearity of spin waves propagating in an yttrium-iron garnet (YIG) ferrite film enables the process. The reservoir performance is evaluated by completing task-independent tests known as short-term memory (STM) and parity-check (PC) tasks. In addition, a numerical model of the MOEO based reservoir is developed. Results of the numerical simulation of the reservoir performance are in good agreement with the experimental data.

## I. Introduction

During the past decade, an important trend in the development of computing devices has been the implementation of artificial neural networks (ANN) as a specialized hardware. This also pertains to a specific case of the recurrent neural network, which is the reservoir network [1]. A reservoir computing (RC) system or device consists of three layers of artificial neurons: an input layer, a hidden layer that has special properties and is called a 'reservoir', and an output layer. One important advantage of this ANN concept is that only one layer of the network is trained. This is the output layer, and it is trained in a simple linear way. This drastically reduces the training time and ensures success of the training procedure. Furthermore, because the reservoir is not trained, the same reservoir can solve multitude of different problems. In addition, it is extremely easy to reconfigure the network to solve a new problem. Thus, the reservoir is capable of multitasking.

Many computational tasks have been successfully solved using the RC model. In particular, this model can be applied to classification, prediction, and generation of time-dependent data structures. Since the reservoir does not require training, it can be viewed as a black box, representing a nonlinear dynamical system, where the design and operating principle are not essential. The only important features of such a system are the ability to map data into a higher dimensional space and the presence of short-term memory. From this perspective, any nonlinear dynamical system with short-term memory (i.e., a temporally non-local response) can serve as a reservoir. This idea led to the concept of physical reservoir computing (PRC) [2,3].

The main characteristics of the physical reservoirs are computational performance, processing speed, memory, power efficiency and scalability of physical systems. The computational performance and the memory are usually evaluated using specially designed benchmark tests that probe both the fading memory and the nonlinear separation ability of the dynamic system. The tests are consistent throughout the literature; therefore, they can be used to compare performance of different PRC concepts. The scalability of a physical reservoir is a property to miniaturize the device-prototype.

In recent years, several hardware platforms have been proposed for physical reservoir computing implementation. The works [4,5] present mechanical systems for the realization of PRC. However, such systems are slow and are unlikely to be implemented as a small chip in the future.

A fiber-optical PRC implementation was proposed in [6-8]. The device represents a system with a delayed feedback. The reservoir represents an optoelectronic loop (or ring). A fiber-optical delay line plays the role of a memory component, and the electro-optical modulator is responsible for the nonlinear mapping and data input. In the modulator, a new input into the reservoir is nonlinearly combined with previous data inputs delayed in time. The optical fiber performs the delay operation – it takes time for the data injected into the fiber to reach its output port. The entire process ensures efficient nonlinear mixing of the inputs. An integrated-optic approach based on utilization of nonlinear microring resonators [9] as physical reservoirs was also reported recently [10,11]

The concept of the magnonic PRC was proposed in the work [12]. The PRC system was in the form of a magnonic active ring resonator based on a YIG film. The ring represents a microwave spin-wave delay line, the input port of which is connected to its output port with a positive feedback loop, similar to the case of the fiber-optical PRC. However, the nonlinear mapping operation is now performed using nonlinearity of the spin-wave dynamics. To convert a magnonic active ring resonator into a PRC device, the presence of a mechanism to inject input data into the reservoir is essential. For this purpose, an electronically controlled attenuator is added to the feedback loop, which is used to control the gain in the ring. Thus, data injection into the reservoir is achieved by introducing a time dependence in the loop gain factor. Subsequent improvements to the concept, as well as advancements building on it, are detailed in Refs. [13-18].

Recently, a combined dynamics and nonlinearity of magnonic-optoelectronic oscillators (MOEOs) attracted attention of researchers [19-21]. Such an active ring system has the advantages of optoelectronic ring, such as a large delay time and low propagation losses of a microwave signal, as well as rich nonlinear dynamics of spin waves. Adding an element for data input to such an active ring resonator turns it into a magnonic-optoelectronic physical reservoir.

In the present work we investigate a magnonic-optoelectronic physical reservoir which was obtained by adding an element for data input to a MOEO. In contrast to works [6-8], the device is designed in such a way that the fiber-optic delay line operates in linear regime providing short-term memory whereas a nonlinearity is provided by spin-wave delay line only.

The paper is organized as follows. Section II presents a model of the MOEO based physical reservoir. Section III describes a characterization of the microwave and optical components utilized for the magnonic-optoelectronic reservoir assembly. Section IV discusses experimental and theoretical results on the reservoir performance. Section V provides summary and conclusions.

## II. A model of magnonic-optoelectronic reservoir

A schematic diagram of the physical reservoir under study is shown in Fig. 1. The reservoir has a ring circuitry consisting of an optical path shown by red color and a microwave path shown by blue color. The optical path of the reservoir consists of a monochromatic optical signal source, two electro-optic modulators of the light intensity, an optical fiber and a fast photodetector. The microwave path comprises a spin-wave delay line, a variable attenuator, two broadband amplifiers, an output directional coupler, as well as the short interconnection cables.

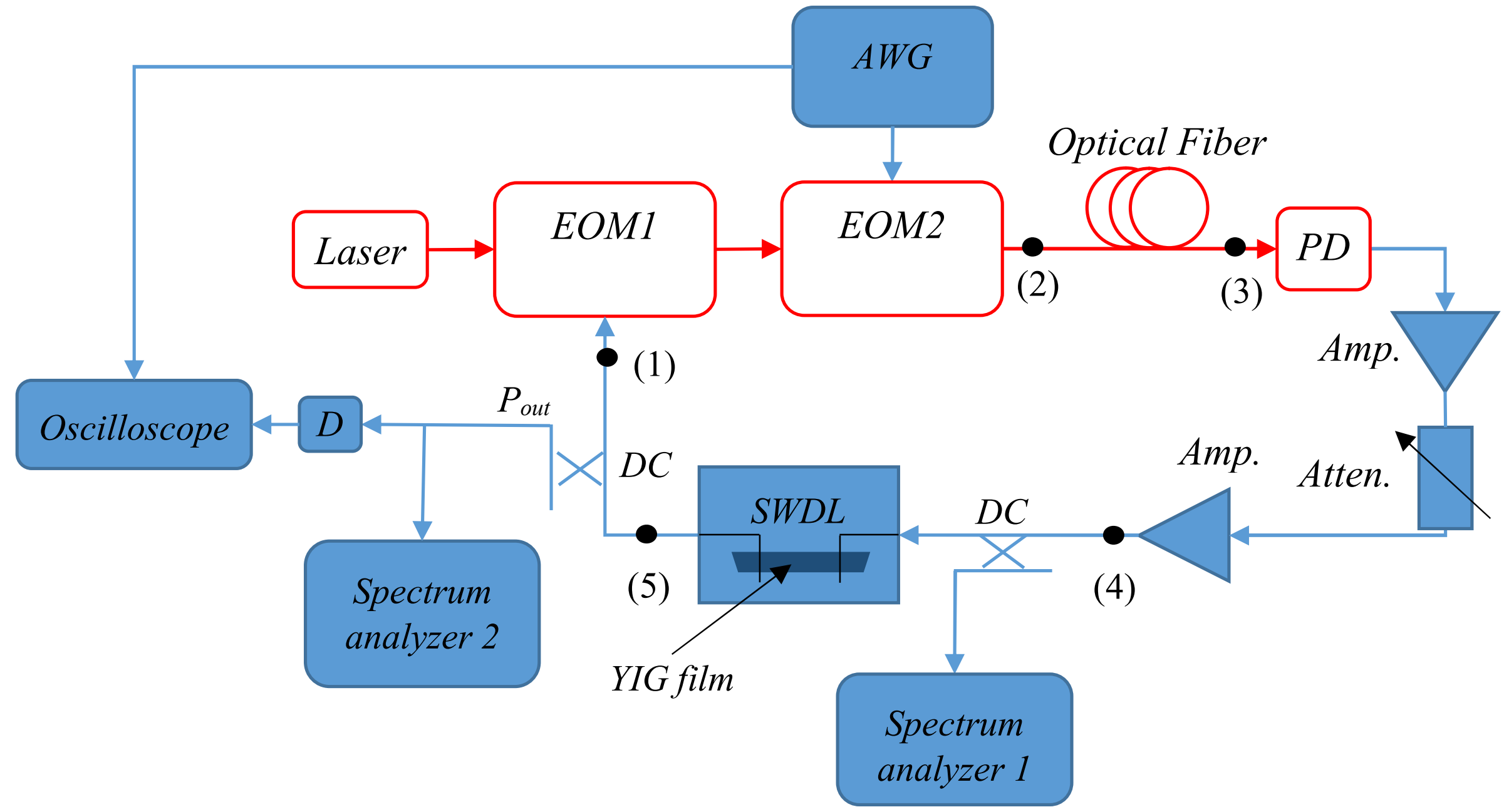

Fig. 1. A schematic diagram of magnonic-optoelectronic physical reservoir (EOM – electro-optical modulator; AWG – arbitrary waveform generator; PD – photodetector; Amp. – amplifier; Atten. – attenuator; DC – directional coupler; SWDL – spin wave delay line; D – detector diode)

The operation of the device is based on nonlinear transient processes between two steady states of self-oscillations distinguished by oscillation power. The transient processes are nonlinear due to the development of four-wave parametric processes in the YIG film utilized in the spin-wave delay line. The optical path operating linearly and is used to introduce additional delay time into the ring. Input data is injected into the reservoir by applying voltage to the second electro-optical modulator (EOM2).

In order to simulate the reservoir operation, we developed a theoretical model. Above the self-oscillation threshold, the ring generates a continuous-wave microwave signal at a particular frequency corresponding to one of the resonant modes of the ring [22]. This frequency satisfies the phase balance condition for the magnonic-optoelectronic ring given by the following expression: $k_{oe}l_{opt}+k_{sw}d+\varphi_e = 2\pi n$, where $k_{oe}$ is a wavenumber of the envelope of the carrier optical wave, $l_{opt}$ is a length of the optical path, $k_{sw}$ is a wavenumber of the spin wave, $d$ is a SW propagation path, $\varphi_e$ is a phase shift of the microwave signal in the remainder of the electronic circuit, and $n$ is any integer.

To simplify the model, we assumed that the phase shift $\varphi_e$ of the microwave signal in the connecting microwave components is negligible compared to the phase accumulation by the spin wave delay line. Therefore, we assume that the full phase shift of the microwave signal in the optoelectronic ring equals $k_{oe}l_{opt}+k_{sw}d$. The frequency $\omega_0$, for which this e phase-balance condition is satisfied, is a resonant frequency of the optoelectronic ring.

Note that the nonlinear spin wave propagating in the delay line accumulates an additional phase shift $\varphi_{NL}$ as a result of four-wave nonlinearity. This effect modifies the phase balance - the balance condition is now satisfied for a different frequency $\omega_1 = \omega_0 + V_g \varphi_{NL}/d$. However, as shown in [15], during the transient process, the nonlinear shift of the auto-oscillation frequency is quite small – it does not exceed hundreds of kHz. In addition, the nonlinear frequency shift does not make a significant contribution to the reservoir operation, since the detector $D$ (Fig. 1) removes information about the microwave phase.

As such, keeping in mind that the waves circulating in the ring add up in phase at the resonant frequency, the problem of modelling self-oscillations in the ring reduces to considering the amplitude balance condition. This condition has the following form: $\left|\dot{H}_{OL}\right| = A_i / A_{i-1} = 1$, where $A_i$ and $A_{i-1}$ are the amplitudes of the microwave signal after the $i$-th and ($i$-1)-th periods of circulation of the signal in the ring. Self-oscillations develop if the open-loop gain is strictly greater than unity $\left|\dot{H}_{OL}\right| > 1$ [15]. With this open-loop gain, the signal amplitude will increase with each circulation around the loop. As the amplitude of the circulating signal increases, the signal loss caused by the nonlinear damping of spin waves in the spin wave delay line increases, and the growth of the signal amplitude slows down. At some point, the loss inserted by the spin-wave delay line will become equal to the gain in the remainder of the system. At this point, the system reaches a dynamic equilibrium characterized by a constant amplitude of the output microwave signal. In this steady state, $|\dot{H}_{OL}|$ equals unity.

Let us consider the power transmission coefficient of the magnonic-optoelectronic open loop. It can be written as:

$$T_{\mathrm{P}} = H_{FODL} G_0 H_{exc} H_{SW}\left(|u|^2\right) H_{rec}\left(1-\kappa_{\mathrm{out}}\right), \tag{1}$$

where $|u|$ is dimensionless spin wave amplitude, $H_{FODL}$ is transmission coefficient of fiber-optic delay line, $G_0$ is amplifier gain; $H_{exc}$ and $H_{rec}$ are transmission coefficients of excitation and reception of spin waves by the transducers (antennas) of the spin-wave delay line, respectively [23], $H_{\mathrm{SW}}(|u|^2)$ is the loss of the signal carried by the excited spin wave in the magnetic film (which also accounts for nonlinear damping of SW, therefore the $|u|^2$ dependence), and $\kappa_{\mathrm{out}}$ is the power coupling coefficient for the directional coupler. Using (1), one can obtain a recurrent expression describing the microwave signal power at the ($i$-1)-th and $i$-th round trip in the ring:

$$P_i = P_{i-1} T_{\mathrm{P}} = H_{FODL} G_0 H_{exc} H_{SW}\left(|u_{i-1}|^2\right) H_{rec}\left(1-\kappa_{\mathrm{out}}\right) P_{i-1}. \tag{2}$$

Similarly, the microwave signal power at the output of the ring can be calculated using the following expression:

$$P_{\mathrm{out},\,i} = H_{FODL} G_0 H_{exc} H_{SW}\left(|u_{i-1}|^2\right) H_{rec} \kappa_{\mathrm{out}} P_{i-1}. \tag{3}$$

The transmission coefficient of the FODL is given by $H_{FODL} = P_{out}^{FODL} / P_{in}^{FODL}$, where $P_{in}^{FODL}$ is microwave-signal power fed into the EOM1 input port, and $P_{out}^{FODL}$ is microwave- signal power exiting the output port of FODL. The latter can be computed using the following expression:

$$P_{out}^{FODL} = R\left(\alpha_{EOM1}\alpha_{EOM2} S P_{las}\, exp\left(-\alpha_{opt} l_{opt}\right) J_1\left(V_0\pi/V_\pi\right) sin\left(V_{b1}\pi/V_\pi\right)\right)^2, \tag{4}$$

where $R$ is the photodetector resistance; $\alpha_{\mathrm{EOM1}}$, $\alpha_{\mathrm{EOM2}}$ are optical losses in EOM1 and EOM2, respectively, $S$ is the sensitivity of photodetector, $P_{las}$ is power of laser optical radiation, $J_1(.)$ is Bessel function of the first kind of order 1, $V_\pi$ is the half-wave voltage of EOM1, $V_0$ is the amplitude of the modulating-signal voltage supplied to the EOM1 input, which is related to $P_{in}^{FODL}$ through the following expression: $V_0 = \sqrt{P_{in}^{FODL} \cdot R_{50}}$, where $R_{50}$ is input impedance of the EOM which is matched to the 50 Ω characteristic impedance of the feeding line.

We computed the transmission coefficient $H_{SW}(|u|^2)$ and the total loss of the microwave signal inserted by the SW delay line using the methods described in [24,25]. This allowed us to carry out numerical simulations of the output signal of the physical reservoir. These results will be shown later on.

### III. Characterization of the components of the physical reservoir

We assembled the experimental prototype of the magnonic-optoelectronic physical reservoir shown in Fig.1 from components with the following characteristics. Two broadband microwave amplifiers used to compensate for losses in the ring have the saturation level of 1 W. Optical radiation power of the laser was 20.2 dBm at a wavelength of 1550 nm. The half-wave voltage for the first electro-optical modulator EOM1 $V_\pi$ = 3.3 V. This device transferred the output microwave signal of the SW delay line to the optical carrier. The role of the second electro-optical modulator EOM2 was to enter input data into the reservoir. This was done by applying a signal from AWG to the control port of the modulator. The half-wave voltage for second modulator $V_\pi$ = 1.8 V. A change in the control voltage modifies the total gain in the ring and, consequently, the amplitude of the circulating microwave signal. The length of the optical fiber was 20 m. The role of directional couplers inserted in front of SWDL and EOM1 was to fan out small fractions of the microwave signal for monitoring the microwave signal with spectrum analyzers. The coupling coefficient for the couplers was 10 dB. This ensured that the SWDL operated in a nonlinear mode, but the optical path in a linear one. In this way, the role of the optical path was just to add extra time delay and to inject input data into the ring. The output signal of the reservoir was first rectified by a microwave diode and then was fed into an oscilloscope for registering the envelope of the output signal.

The spin-wave delay line represented a 10.1-μm-thick, 2-mm-wide and 25-mm-long strip of a single-crystal YIG film grown with liquid-phase epitaxy on gadolinium-gallium garnet substrate. The saturation magnetization of the film was 1750 G. An external bias magnetic field of 1318 Oe was applied to the film in its plane and perpendicular to its long side. This field orientation corresponds to the magnetostatic surface spin wave (MSSW) configuration. The distance between the input and the output antennas of the delay line was 4.7 mm. The magnetic loss parameter for the film $\Delta H = 0.5$ Oe.

A characterization of the reservoir components was carried out in several stages. In the first stage, the performance characteristics of the SW delay line were measured with a Rohde&Schwarz ZVA40 vector network analyzer. Firstly, the linear mode of its operation was investigated. To this end, microwave power of -10 dBm was applied to its input port. Fig. 2(a) shows the amplitude-frequency characteristics of the delay line recoded in this mode. One can see that the insertion loss is minimum at a frequency of 5.719 GHz and that the respective minimum loss amounts to –7.5 dB. The last figure is very important because the ring is expected to self-oscillate in the vicinity of this frequency. Fig. 2(b) presents the experimental dispersion characteristics (dots in the figure) obtained by recording the phase-frequency characteristic of the delay line with the vector analyzer data. The continuous line in the figure is for the theoretical dispersion characteristics for Damon-Eschbach MSSW mode (see e.g. [26]). Ones sees an excellent agreement between the experiment and the theory.

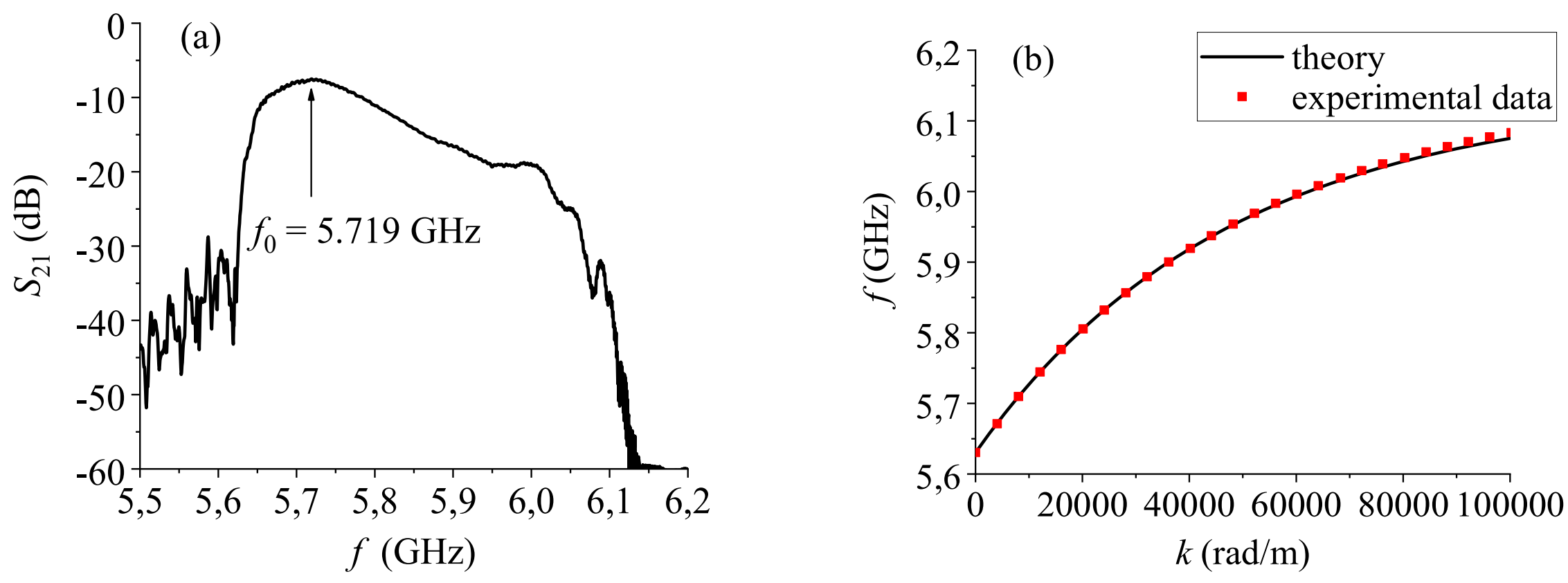

Fig. 2. (a) Amplitude–frequency characteristic of the spin-wave delay line. (b) Experimental (red symbols) and theoretical (black solid line) dispersion characteristics of the spin waves propagating in the YIG film.

The next step was to investigate the operation of the SW delay line in a large-amplitude (nonlinear) mode. Fig. 3(a) shows the insertion loss as a function of the input microwave power measured at a frequency of 5.719 GHz. One can see that for powers above 6 dBm, the inserted loss starts to grow due to the onset of nonlinear damping of the spin waves. The corresponding transmission characteristic is shown in Fig. 3(b). Ones sees a departure from linear scaling of the output power with the input one. The departure starts at input power of 4 mW (which roughly corresponds to 6 dBm) and slowly leads to signal saturation with a further increase in input power.

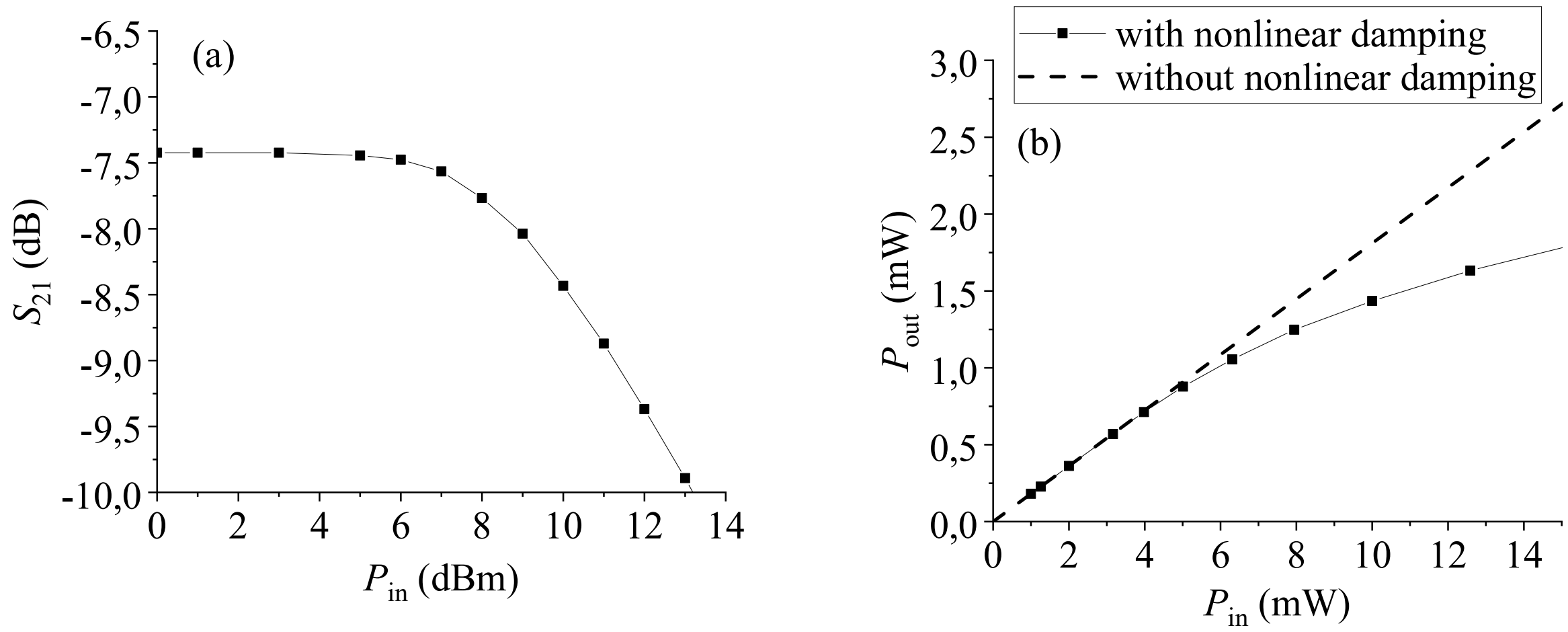

Fig. 3. Insertion loss (a) and output microwave power (b) as a function of input microwave power of SW delay line measured for 5.719 GHz.

The second stage of the ring characterization was an investigation of the optical path. The first step was to take the transmission characteristic of the first electro-optical modulator using a setup shown schematically in Fig. 4(a). To enable this, we kept the laser output at 20.2 dBm, and measured intensity of light exiting the electro-optical modulator for different values of control voltage $V_{b1}$. Fig. 4(b) presents this result. The symbols in the figure are the experimental data. The solid line shows the approximating function $A(1+\sin(bV_{b1}+c))$, where

$A=\alpha_{EOM1}[arb.\,u.]\cdot P_{las}[mW]/2=11.02\,mW,\quad b=\pi/V_{\pi}=1.039\,V^{-1}, c=-0.3\,rad$.

From these data, we find that the optical loss inserted by the modulator and at the fiber optic connections is $\alpha_{EOM1}$ = –6.56 dB and the modulator quadrature points correspond to $V_{b1}$ = 0.25, 3.38 and 6.3 V.

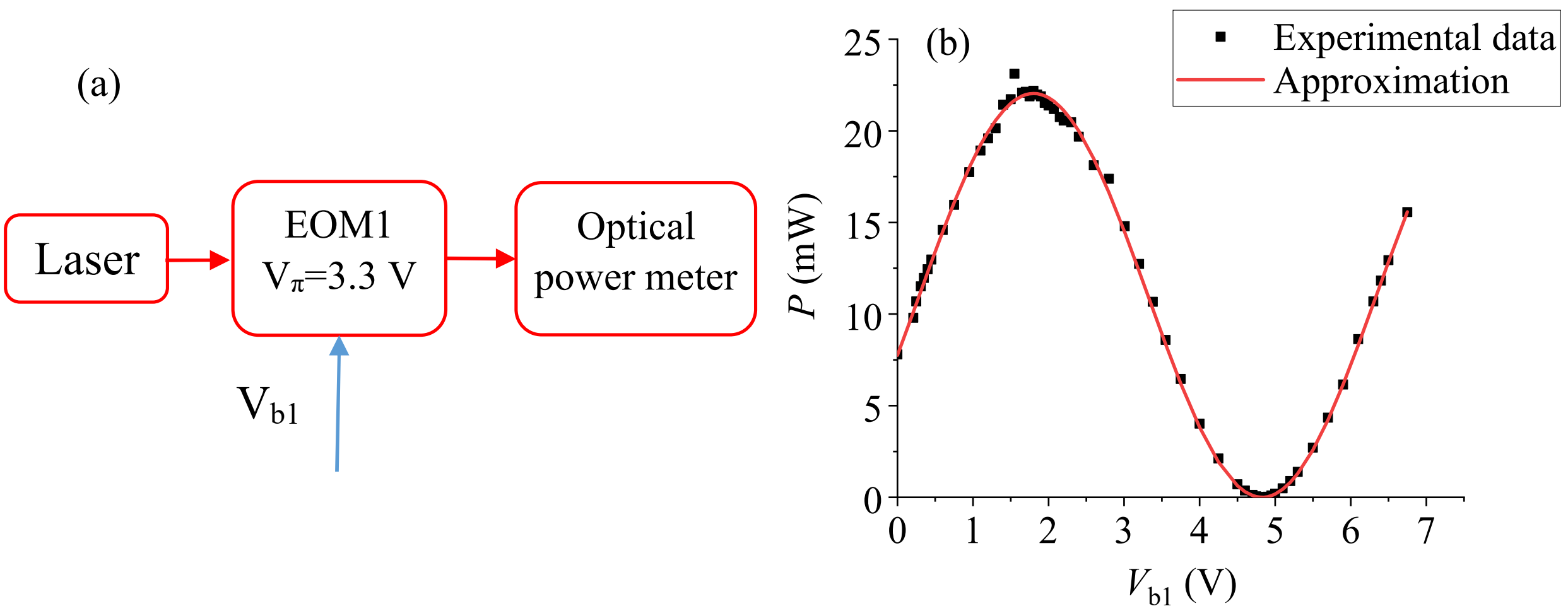

Fig. 4. (a) Schematic diagram of the setup for measuring the transmission characteristic of the electro-optical modulator EOM1. (b) The transmission characteristic of the electro-optical modulator EOM1.

The next step was to take the transmission characteristics of the cascade connection of EOM1 and EOM2. The measurement setup employed to this end is shown in Fig. 5(a). The operating point of the first modulator was set to quadrature $V_{b1}$ = 0.25 V. The recorded trace is shown in Fig. 5(b). The function $A(1+\cos(bV_{b2}+c))$, where $A = \alpha_{\text{EOM1}}[arb.\ u.]\alpha_{\text{EOM2}}[arb.\ u.]P_{las}[mW]/2 = 1.143\ mW,\ b = \dfrac{\pi}{V_\pi} = 1.826\ V^{-1}$, $c = 0.1921\ rad$ fits it quite well. This result permits extraction of optical loss inserted by the second modulator. We found that losses at the fiber optic connections are $\alpha_{\text{EOM2}} = -7$ dB.

In the following, we will use the second modulator to inject input data into the reservoir. This is implemented by varying the total insertion loss of FODL in time. The operating point of EOM2 is then set to near the maximum of the transmission characteristic at $V_{b2} = 0$ V (see Fig. 5(b)).

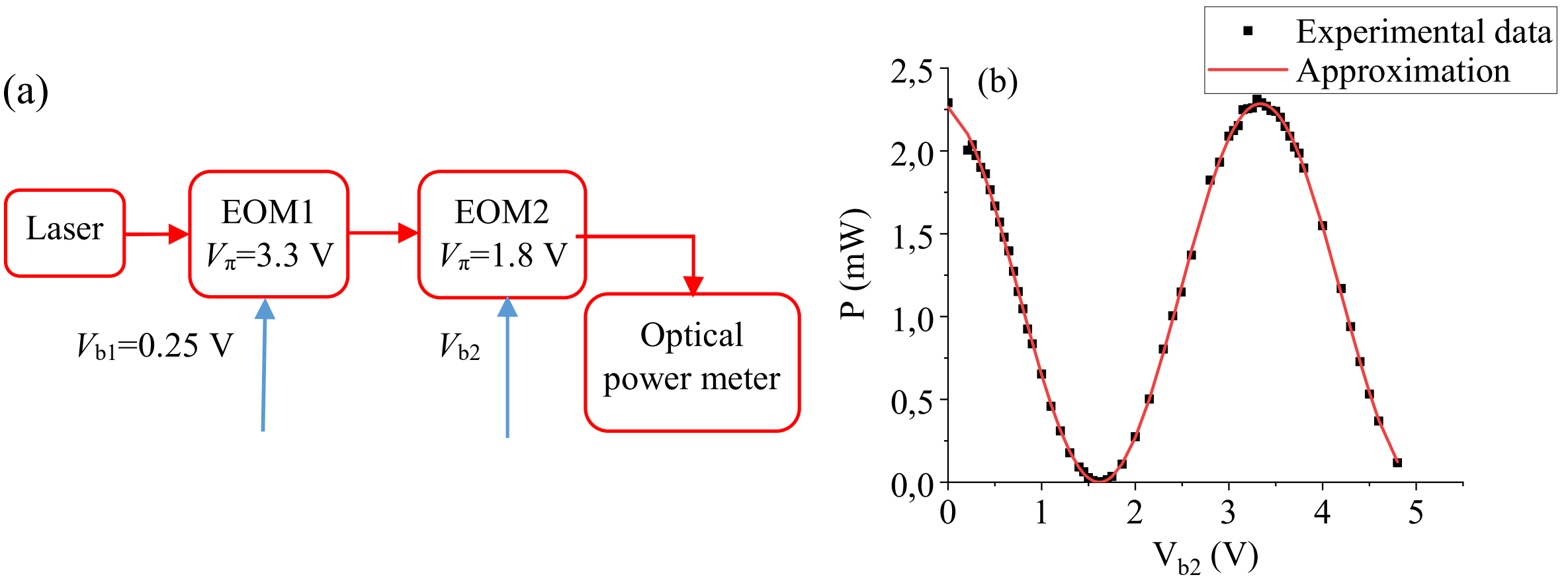

Fig. 5. (a) A diagram of the setup for measuring the transmission characteristic of the electro-optical modulator EOM2. (b) Optical power at the output of the electro-optical modulator EOM2 as a function of the bias voltage $V_{b2}$.

The last step was to determine the linear dynamic range of the FODL. First we measured the FODL insertion loss as a function of control voltage $V_{b2}$ (see Fig. 6) for the microwave signal frequency of 5.719 GHz and power of 0 dBm. Knowing this characteristic is needed to numerically model the physical reservoir output. This value underlies the magnitude of change in the transmission coefficient of the FODL when input data is injected into the reservoir. Then we determined the threshold level for the nonlinear compression of the microwave signal transmitted by FODL as an envelope of the carrier optical wave. Fig. 7 shows the insertion loss and the output microwave power as a function of the input microwave power fed into the EOM1 of the FODL. One can see that output power saturates above 3.2 mW of input power. Note, that the effect of saturation is due to the nonlinear character of the transmission characteristics of EOM1.

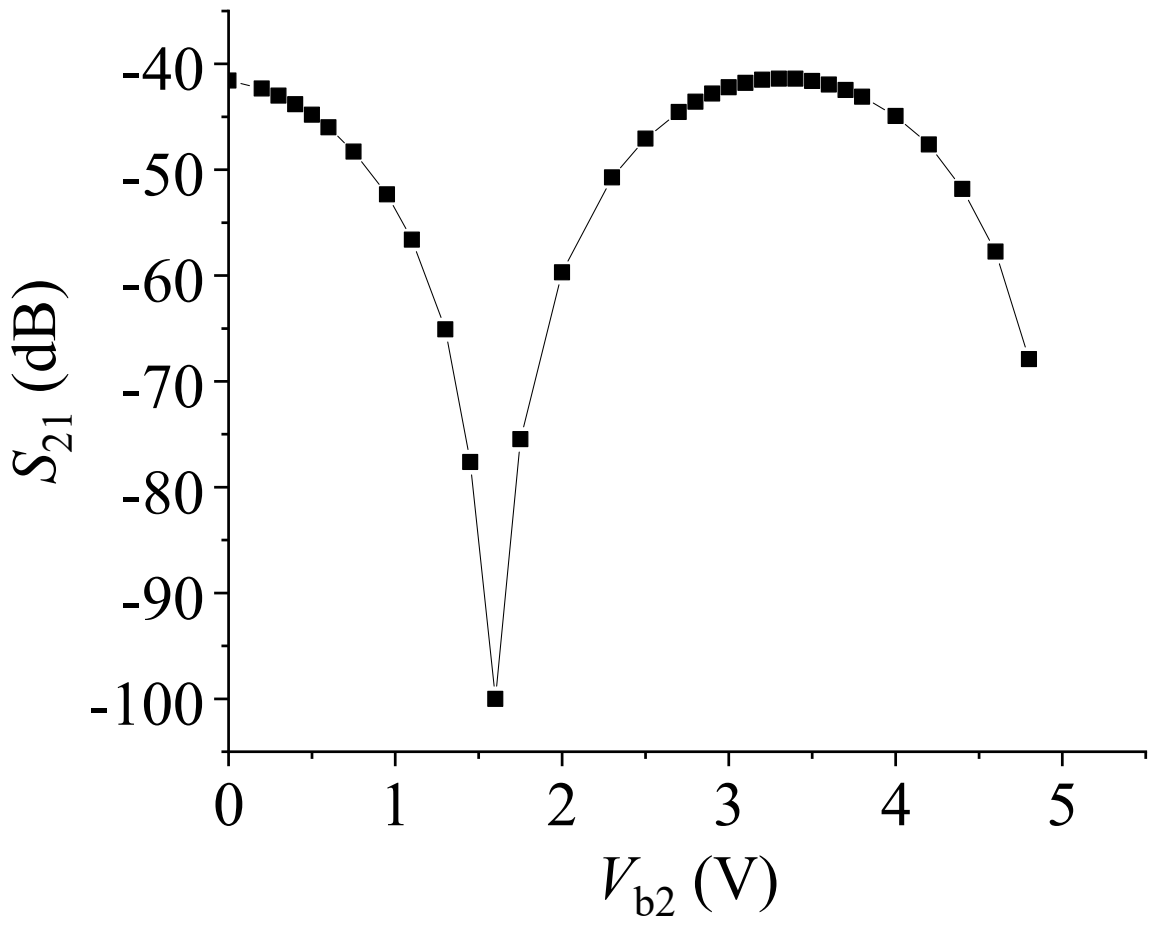

Fig. 6. Transmission coefficient of FODL as a function of $V_{b2}$ measured for frequency 5.719 GHz.

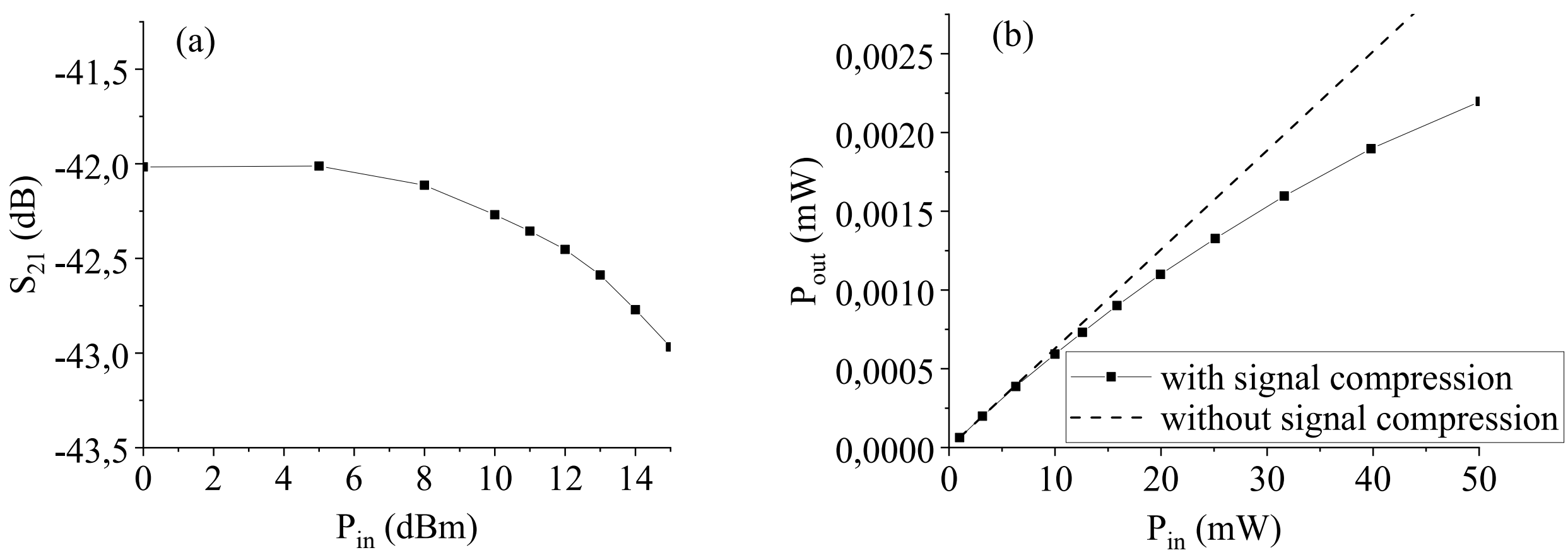

Fig. 7. Insertion loss (a) and output microwave power (b) as a function of input microwave power of the FODL measured for 5.719 GHz.

## IV. Investigation of the reservoir performance

As is mentioned above, the SW delay line must operate in a nonlinear mode, whereas the FODL in a linear one. After the onset of a self-oscillation in the ring, this condition can be satisfied if a microwave signal entering the SW delay has power exceeding 6 dBm, and power of a microwave signal fed into the control port of the EOM1 is below 5 dBm. For our experimental reservoir, we find that for the total ring gain of $G_0$ = 0.05 dB power of the microwave input signal of the SW delay line is about 6.4 dBm, and power of the microwave signal fed to the EOM1 is –8 dBm. These values were monitored with the respective spectrum analyzers as shown in Fig. 1. For $G_0$ > 0.09 dB the ring was abruptly switching from the single-frequency self-oscillation mode to frequency-comb generation.

We started with investigating the transient processes in the ring. To this end, we applied a periodic sequence of rectangular pulses with a duration of 250 μs to the control port of EOM2. This time interval was sufficient for completing the transition between different steady-state self-oscillation levels of power of the microwave signal circulating in the ring. The control-pulse amplitudes $V_{EOM2}$ were 10, 15 and 25 mV, they were reducing the ring gain $G$ by 0.04 dB, 0.06 dB and 0.09 dB, respectively.

Representative output signals of the reservoir are shown in Fig. 7. One can see that in the absence of input data, the ring operates in the self-oscillation mode. For $V_{EOM2}$ = 25 mV, the ring leaves the self-oscillation mode for the duration of the control pulse, because the open-loop gain becomes less than unity (see Fig. 8(a)). For $V_{EOM2}$ = 15 mV, the steady self-oscillation also stops for the pulse duration (see Fig. 8(b)). For $V_{EOM2}$ = 10 mV, the self-oscillation does not stop, but a transient process takes place that gradually changes the level of oscillating power. The respective oscillation-amplitude rise and fall times were found $\tau_{rise} \approx 10$ μs and $\tau_{fall} \approx 38$ μs respectively. This ring operation mode complies with the operating principle for a self-oscillator-based physical reservoir.

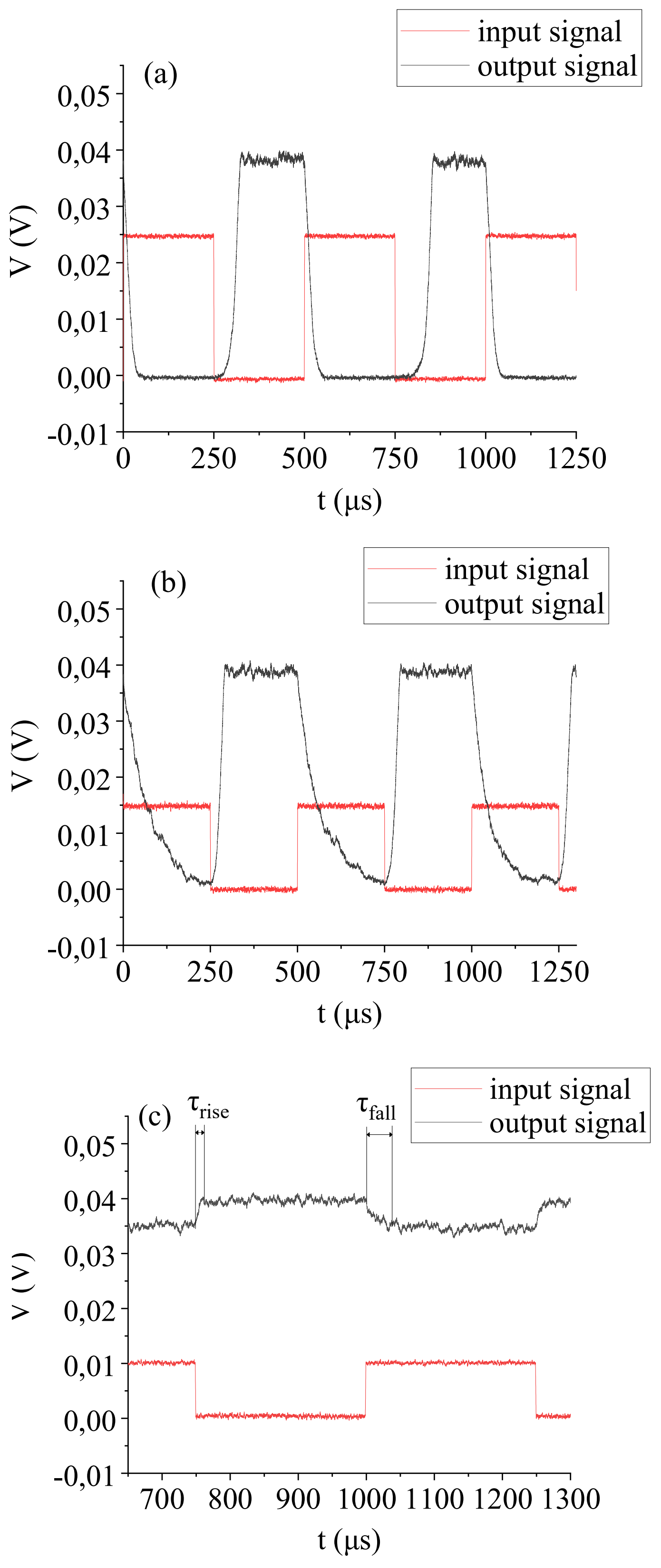

Fig. 8. – Fragments of oscilloscope traces of the reservoir output signal when applying periodic voltage pulses with an amplitude of 25 mV (a), 15 mV (b), 10 mV (c)

These experimental results are in good qualitative agreement with our numerical simulations (Fig. 9). This confirms that he developed model adequately describes the reservoir dynamics.

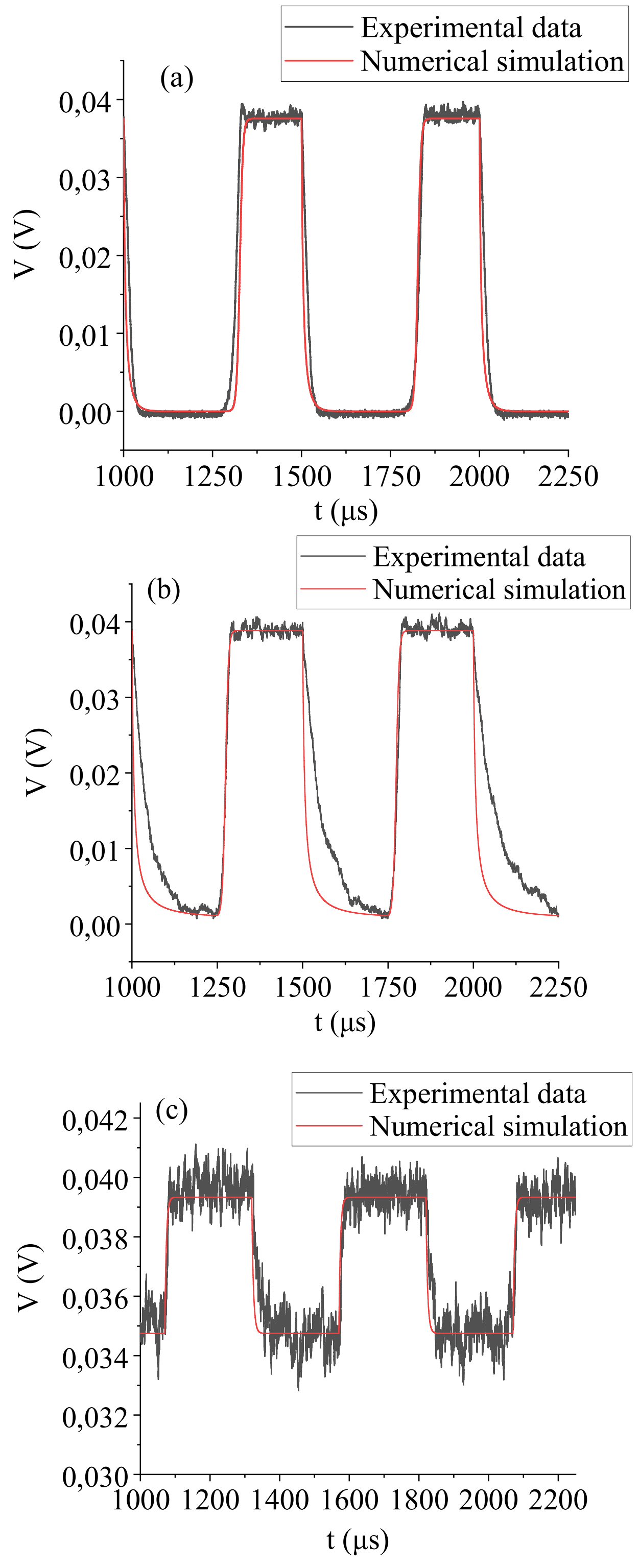

Fig. 9. Measured and simulated waveforms from the reservoir output for a periodic sequence of pulses with a duration of 250 μs and an amplitude of 25 mV (a), 15 mV (b), 10 mV (c)

Turn to the experimental results. The experimental conditions were as follows. A random binary sequence was injected to the reservoir using EOM2. A duration of the pulses was 5, 10, 20, 50 μs and the peak voltage of the pulses in this sequence was 15 mV. Note that the gain $G_0$ was slightly increased to a value of $G_0 = 0.07$ dB. This was done so that when a control pulse was applied, the reservoir would remain in the auto-oscillation regime. Fig. 10 shows fragments of typical oscilloscope traces of the input and output signals of the reservoir for pulse durations of 10 and 50 μs.

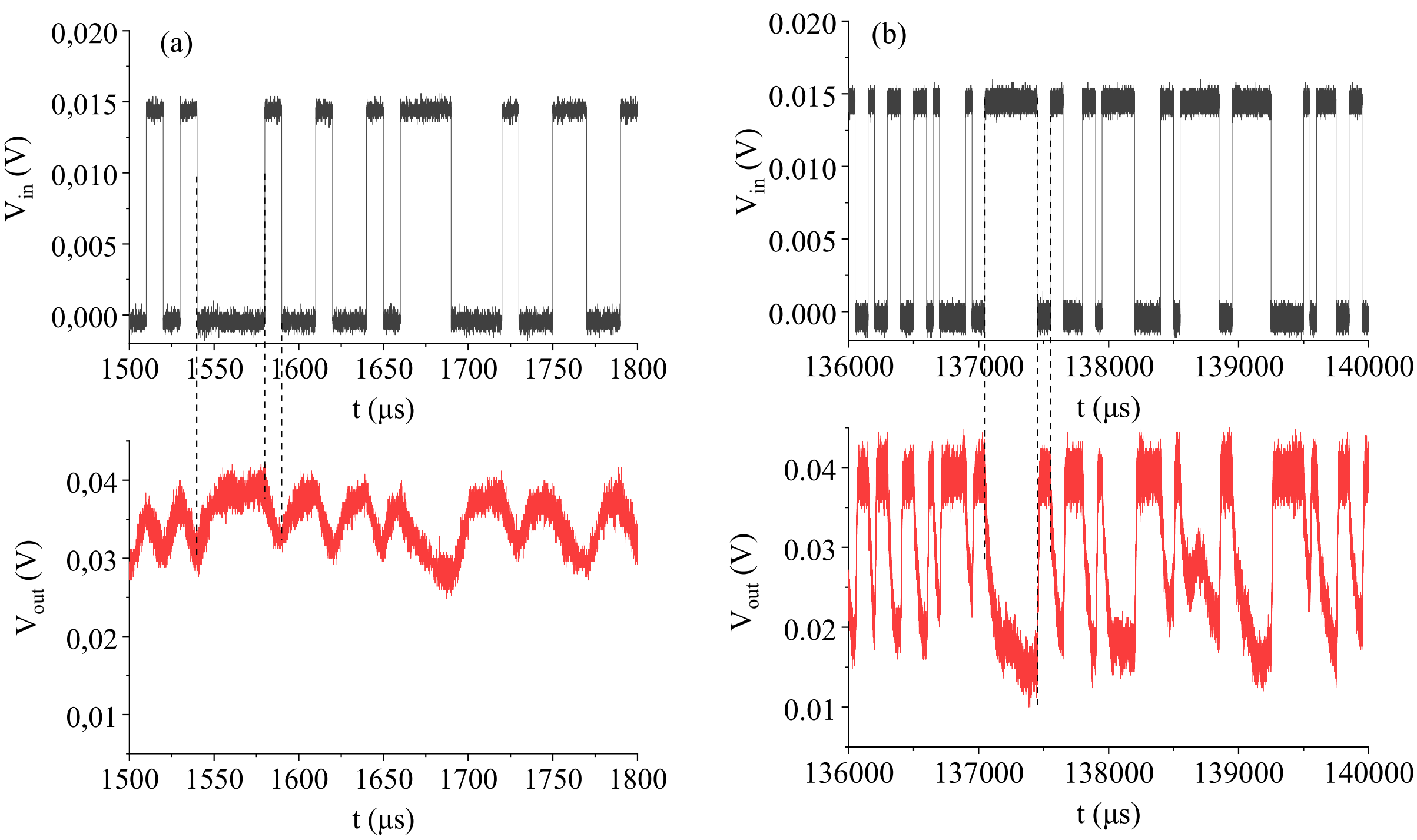

Fig. 10. The binary data input sequence (upper part) and the respective waveform from the reservoir output (bottom part) for pulse durations of 10 μs (a) and 50 μs (b)

The process of testing the magnonic-optoelectronic reservoir is similar to that described in [6]. We use a random binary sequence consisting of 4200 pulses for testing. These are the input signal pulses applied to the control port of EOM2 from the AWG. The respective 4200 outputs, each consisting of multiple samples collected over each input interval, were recorded using an oscilloscope. In order to eliminate the effect of initial conditions on the system dynamics, the first 200 outputs are discarded. The next 2000 outputs are used for the training process. The remaining 2000 outputs are used for testing. For testing, we use two well-established benchmark tasks, the short-term memory (STM) and parity-check (PC) tests. They characterize the reservoir performance in the so-called "task-independent way".

From Fig. 11, it can be seen that the maximum value of the STM capacity of 1.08 is achieved with a pulse duration of 20 μs. An increase of the pulse duration to 50 μs leads to a decrease in the STM capacity to 1.04. An increase of the pulse duration leads to an increase of the PC capacity. The maximum PC capacity was 0.74 with a pulse duration of 50 μs.

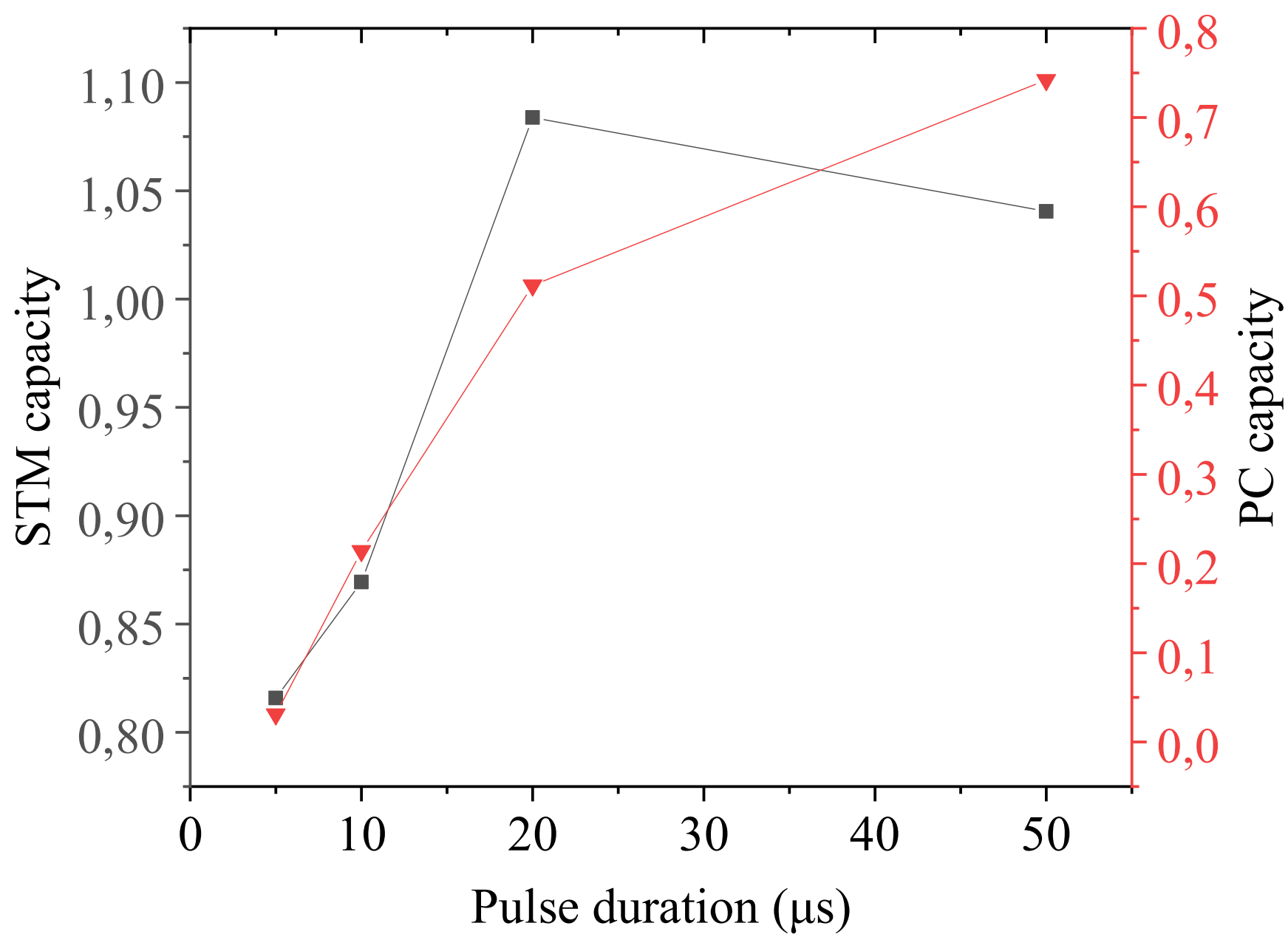

Fig. 11. – STM (black squares) and PC(red triangles) capacities as a function of input pulse duration.

Finally, we simulated numerically the output signal of the reservoir. The calculations were carried out using the parameters of the experimental setup. Fig. 12 shows the results of numerical simulation and experimental measurements of the reservoir output signal. It can be seen that the theoretical and experimental results are in good agreement.

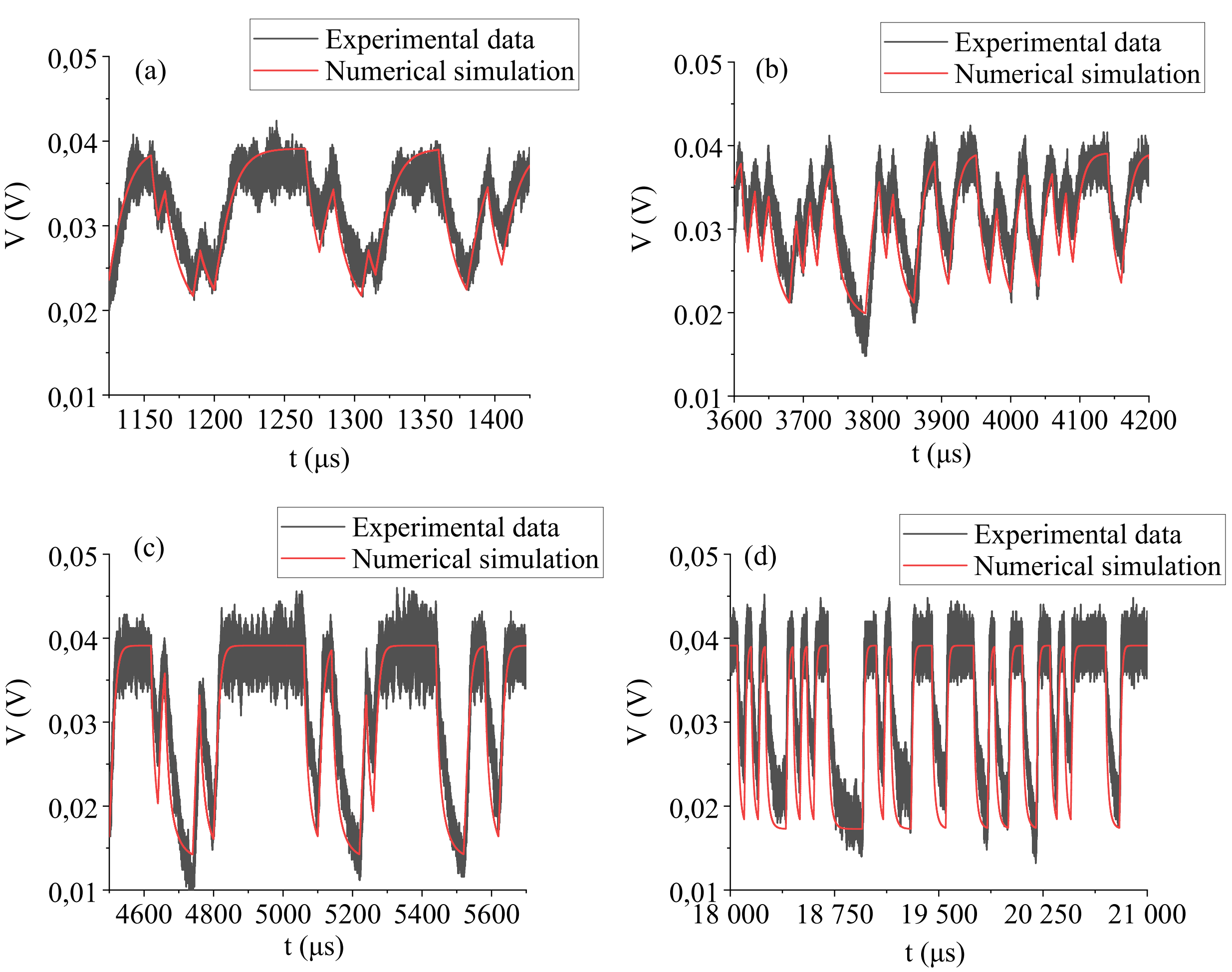

Fig. 12. Measured and simulated waveforms from the reservoir output for a random sequence of pulses with a duration 5 μs (a), 10 μs (b), 20 μs (c), 50 μs (d).

## V. Conclusion

We have proposed and developed a magnonic-optoelectronic reservoir. The device has a structure of an optoelectronic oscillator with a spin-wave delay line as its core element. The optical path of the reservoir operates in the linear mode and is used for providing additional delay time and for entering input data into the reservoir. The spin wave delay line operates in a nonlinear mode and is used for nonlinear mapping operation. The performance of the reservoir has been investigated both experimentally and theoretically. The obtained results are important for further development of reservoirs for physical reservoir computing.

## Acknowledgement

The work at St. Petersburg Electrotechnical University was supported by the Ministry of Science and Higher Education of the Russian Federation (grant number No. FSEE-2025-0008